\newcommand{\remove}[1]{}

\title{A Preliminary Proposal for an Analytical Model for Evaluating the Impact on Performance of Data Access Patterns in Transaction Execution}

\author{Pierangelo Di Sanzo}
\documentclass[12pt]{article}
\usepackage{graphicx}
\usepackage{amsmath}
\date{}

\begin{document}
\maketitle

\begin{abstract}
We present a preliminary proposal for an analytical model for evaluating the impact on performance of data access patterns in concurrent transaction execution. We consider the case of concurrency control protocols that use locking to ensure isolation in the execution of transactions. We analyse scenarios where transactions access one or more sets of data items in the same order or in different order. 
\end{abstract}

\section{Introduction}
\label{introduction}

Transactional systems, like database systems and transactional memory \cite{Thomasian_1998, 10.5555/1855056}, use Concurrency Control Protocols (CCPs) to ensure isolated execution of concurrent transactions. Various CCPs use locking as a basic mechanism to avoid conflicts when transactions access shared data items. Two-phase locking (2PL) \cite{Lausen2009} is a commonly adopted lock acquisition policy that guarantees serializability. In 2PL, locks are acquired and released in two phases. During the first phase, the \textit{acquisition} phase, locks are incrementally acquired with no lock release. During the second phase, the \textit{release} phase, locks are released and no lock are acquired.

We present a preliminary proposal of an analytical model for evaluating the impact on performance of transaction data access patterns for CCPs that use 2PL. In particular, we consider the Encounter Time Locking (ETL) acquisition strategy \cite{10.1145/1345206.1345241}, i.e. locks are acquired during the transaction execution as soon as each data item is accessed, and are released at the end of the transaction after is has been committed. In the case of lock conflict, the transaction is aborted, all locks are released and the transaction is restarted. We note that, due to abort on lock conflict, deadlock can not occur.
We consider different kinds of data access patterns where transactions access one or more sets of data items in the same order or in different order, and we propose a modelling approach for quantify the effects of locking on performance.

\section{Analysed  data access patterns}
\label{scenarios}

We consider two different scenarios. In the first one, that we identify as \textbf{Scenario 1}, we consider transactions that access different data tables. We analyse the case in which transactions access data tables according to different orders, and the case in which transactions access data tables according to the same order. Specifically, in the first case, that we identify as \textbf{Case 1.1}, we consider that a subset of the concurrent threads of the application execute transactions that access data tables in the same order, and another subset of threads execute transactions that access data tables in the inverse order. For example, each transaction executed by the first set of threads accesses data table $A$ and then access data table $B$, while each transaction executed by the second set of threads accesses data table $B$ and then accesses data table $A$. As for the second case, that we identify as \textbf{Case 1.2}, all threads execute transactions that access data tables according to the same order. For example, each transaction accesses data table $A$ and then data table $B$. 
In the second scenario, that we identify as \textbf{Scenario 2}, we consider the following two cases. In the first case, \textbf{Case 2.1}, all transactions access data items in single table in a random way. For example, if there are $10$ data items in the table and all transactions access $3$ data items, a transaction could access in sequence the data items $7,1,4$ and another transaction could access in sequence the data items $10,2,7$. In the second case, \textbf{Case 2.2}, all transactions access data items in the data table according to the same ordering rule. For example, in this case the previous sequences of data items accessed by transactions would be $1,4,7$ and $2,7,10$, respectively.

We analysed the above-mentioned scenarios with real transactional systems that use 2PL. We observed that, both in the case of transactions accessing different data tables and in the case of transaction accessing data items in a single data table, the average transaction execution time decreases by ordering data accesses according to the same order rule. Thus, to investigate this phenomenon from the perspective of the impact of locking, we built an analytical model that captures the effects on the transaction execution time due to the concurrency control protocol depending on the order of accesses to data table and data items. The analytical model we propose is useful to study the phenomena excluding the effects due to shared hardware, like possible effects due to cache memory, or to asymmetric memory accesses in NUMA architectures. In this way, it is possible to isolate and quantify the effect of the concurrency control protocol.

\section{The Analytical model}
\label{mathematical_model}

The approach that we use to build the analytical model is based on the methodology presented in some previous studies on concurrence control protocols in transactional systems, in particular databases \cite{DiSanzoEtAl_MASCOTS2000,DisanzoEtAl_WOSP2010} and transactional memories \cite{5581581, DiSanzo:2012:AMC:2181299.2181585}. In the cited articles, the methodology has been validated using various kind of workloads, and has been demonstrated to be effective for modelling various kind of concurrency control protocols. In this article, we use the same methodology, and we extend the  model for capturing the effects of transaction data access patterns with the ETL acquisition policy.

\subsection{System model and transaction execution model}

We consider a system with $m$ concurrent threads which execute transactions. A transaction can execute transactional operations and code blocks composed of an arbitrary number of (non transactional) instructions. Each transactional operation accesses a data item in a set of $d$ shared data items. Upon each transactional operation, the transaction tries to acquire a lock on the accessed data item. In the case of conflict with a lock already acquired by another transaction on the same data item, the transaction is aborted and restarts. If no conflicts occur along the execution of all transactional operations, the transaction commits and releases all locks. 

\begin{figure}[ht]
	\centering
	\includegraphics[width=0.5\linewidth]{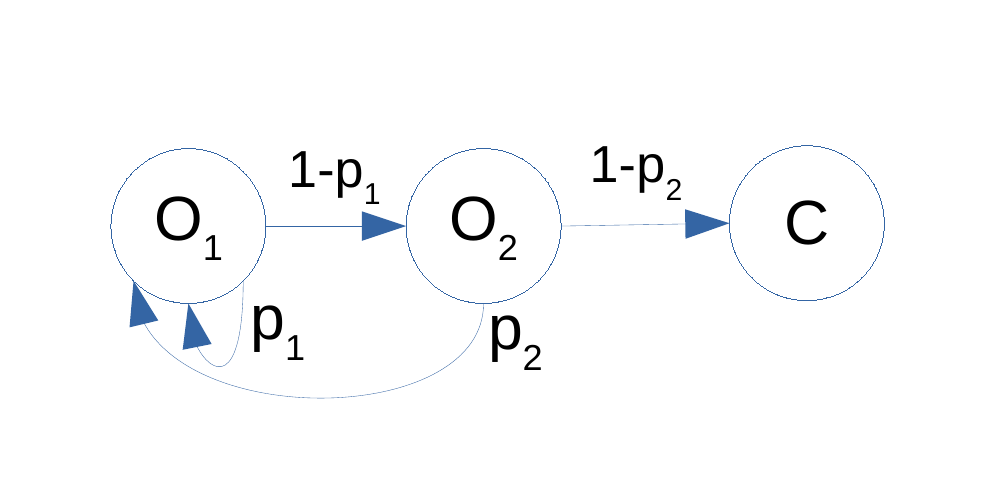}
	\caption{DTMC with 2 transactional operations.}
	\label{SimpleChain2Writes}
\end{figure}

We use a DTMC to model the execution of a transaction with $n$ transactional operations alternated by code blocks. In Figure \ref{SimpleChain2Writes}, we show an example of DTMC for a transaction with 2 transactional operations. The transaction starts with a begin state, denoted as $B$, then executes a code block in state $C1$. Then it executes the first transactional operation in state $O_1$. In the case of conflict upon the transactional operation, the transaction gets aborted and restarts from state $B$. $P_1$ represents the conflict probability. In the case no conflict occurs, it executes the subsequent code block in state $C2$. Then it executes the second transactional operation in state $O_2$. Again, in the case of conflict it gets aborted and restarts from state $B$. The conflict probability is denoted as $P_2$. Otherwise, it executes the last code block in state $C3$ and then commits in state $C$. 

\begin{figure}[ht]
	\centering
	\includegraphics[width=0.9\linewidth]{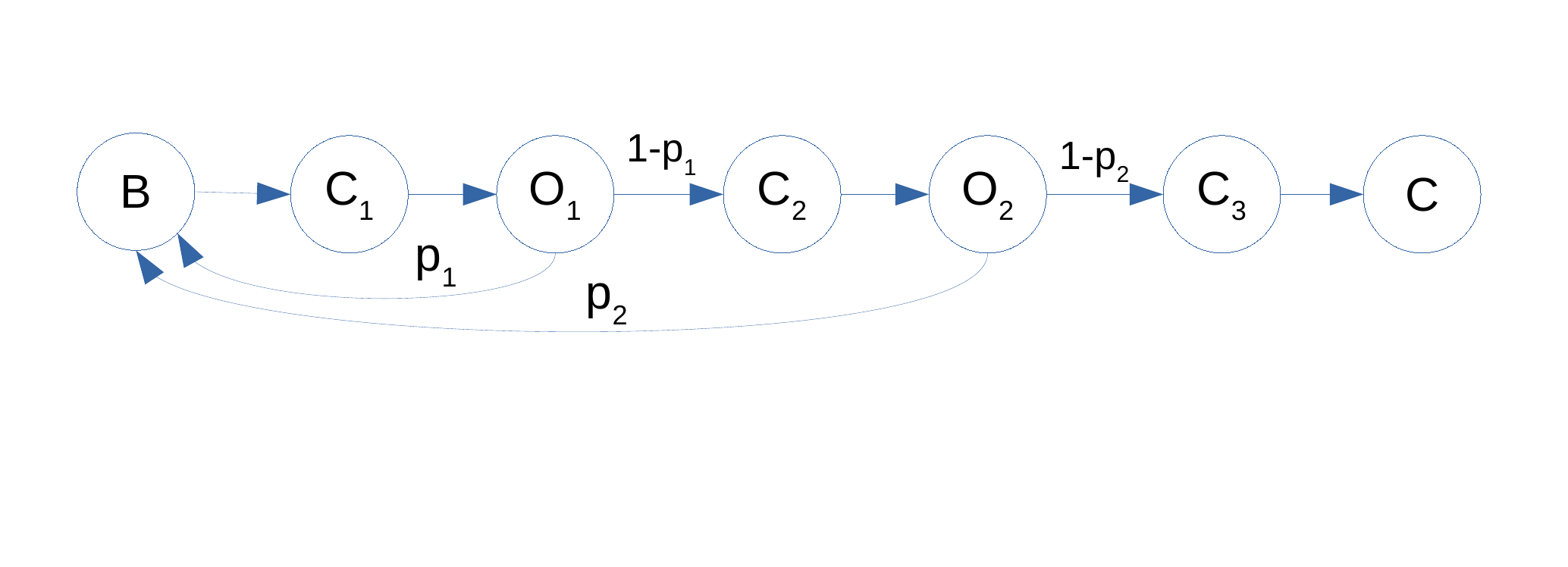}
	\caption{A representation of DTMC with 2 transactional operations.}
	\label{Chain1}
\end{figure}

The above described model is general and can represent transactions with arbitrary sequences of transactional operations and code blocks with arbitrary duration. However, to the aim of our study, it can be further simplified without loss of generality. Indeed, the CTMC is composed of a single absorbing state $C$, and all the others are transient states. Since we are interested in the evaluation of the average number of visits at each state before the transaction commits, we note that some transient states receive the same number of visits as their successors. Thus, they can be incorporated in them. Specifically, $B$  and $C_1$ can be incorporated within the subsequent state $O_1$, all states $C_i$ with $1<i<n$, can be incorporated within each subsequent state $O_i$, and state $C_n$ can be incorporated in state $C$. The new DTMC for a transactions with 2 transactional operations is shown in Figure \ref{SimpleChain2Writes}.

\noindent
In this new transaction model, we assume that $T$ is the vector, with size $n$, of the residence times in states $O_i$, with $1 \leq i \leq n$, and $t_C$ is the residence time in state $C$.

\subsection{Average number of visits at each state}
The transition matrix of the DTMC of a transaction with $n=2$ is the following:

\begin{equation}
P = 
 \begin{pmatrix}
  p_1 & 1-p_1 & 0 \\
  p_2 & 0 & 1-p_2 \\
  0 & 0 & 1
 \end{pmatrix}.
\end{equation}

\noindent
Denoting with $Q$ the transition matrix among the transient states, which in our case corresponds to the matrix composed of the first $2$ rows and columns of $P$, we have

\begin{equation}
Q = 
 \begin{pmatrix}
  p_1 & 1-p_1\\
  p_2 & 0
 \end{pmatrix},
\end{equation}

\noindent
Thus, the fundamental matrix is

\begin{equation}
N = (I-Q)^{-1} =
 \begin{pmatrix}
 \frac{1}{(p1 - 1)(p2 - 1)} & -\frac{1}{p2-1} \\
\frac{p2}{(p1 - 1)(p2 - 1)} & -\frac{1}{p2-1} 
\end{pmatrix}.
\end{equation}

We remark that the element $N_{i,j}$ represents the average number of visits to state $j$ before absorption when the initial state is $i$. Since the initial state of a transaction execution is always $O_1$, then the average number of visits to each state is expressed by the first row of $N$, that we denote as $N_1$. Thus we have:

 \begin{equation}
N_1 =
 \begin{pmatrix}
 \frac{1}{(p1 - 1)(p2 - 1)}, -\frac{1}{p2-1}
\end{pmatrix}.
\label{N_1_2}
\end{equation}

\begin{figure}[ht]
	\centering
	\includegraphics[width=0.7\linewidth]{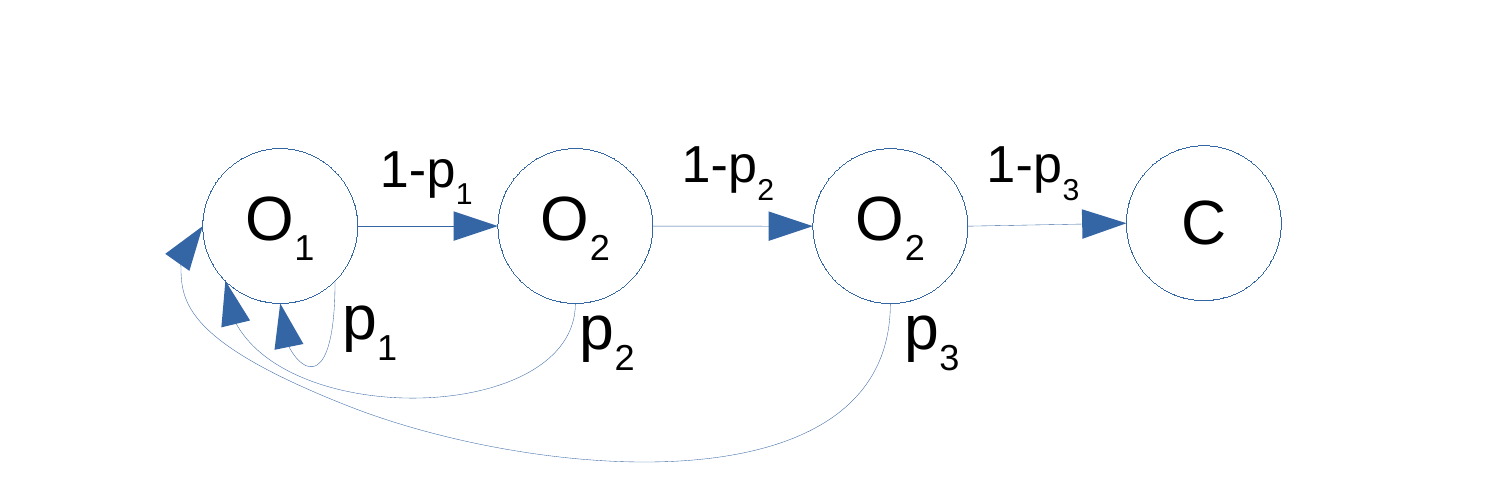}
	\caption{DTMC with 3 transactional operations.}
	\label{SimpleChain3Writes}
\end{figure}

Now we consider the DTMC of a transaction with $n=3$, which we show in Figure \ref{SimpleChain3Writes}.

\noindent
In this case we have:

\begin{equation}
P = 
 \begin{pmatrix}
  p_1 & 1-p_1 & 0 & 0\\
  p_2 & 0 & 1-p_2 & 0\\
  p_3 & 0 & 0 & 1-p_3\\
  0 & 0 & 0 & 1
 \end{pmatrix},
\end{equation},

\noindent
and

\begin{equation}
Q = 
 \begin{pmatrix}
  p_1 & 1-p_1 & 0\\
  p_2 & 0 & 1-p_2\\
  p_3 & 0 & 0 &
 \end{pmatrix}.
\end{equation}

\noindent
Thus, we again calculate $N_1$, achieving

\begin{equation}
N_1 =  
 \begin{pmatrix}
-\frac{1}{((p1 - 1) (p2 - 1) (p3 - 1))} , \frac{1}{(p2 - 1) (p3 - 1)} , -\frac{1}{p3-1} 
\end{pmatrix}.
\label{N_1_3}
\end{equation}

\noindent
Considering the DTMC of a transaction with $n=4$, we have

\begin{equation}
P = 
 \begin{pmatrix}
  p_1 & 1-p_1 & 0 & 0 & 0\\
  p_2 & 0 & 1-p_2 & 0 & 0\\
  p_3 & 0 & 0 & 1-p_3 & 0\\
  p_4 & 0 & 0 & 0 & 1-p_4\\
  0 & 0 & 0 & 0 & 1
 \end{pmatrix},
\end{equation}

\noindent
and

\begin{equation}
Q = 
 \begin{pmatrix}
  p_1 & 1-p_1 & 0 & 0\\
  p_2 & 0 & 1-p_2 & 0\\
  p_3 & 0 & 0 & 1-p_3 \\
  p_4 & 0 & 0 & 0  
 \end{pmatrix}.
\end{equation}

\noindent
Finally,

\begin{equation}
N_1 =
\begin{pmatrix}
\frac{1}{(p_1 - 1) (p_2 - 1) (p_3 - 1) (p_4 - 1)} , -\frac{1}{(p_2 - 1) (p_3 - 1) (p_4 - 1)} , \frac{1}{(p_3 - 1) (p_4 - 1)} , -\frac{1}{p4-1} 
\end{pmatrix}.
\label{N_1_4}
\end{equation}

\noindent
Thus, we can continue to calculate $N_1$ while increasing $n$. However, by observing equations \ref{N_1_2}, \ref{N_1_3} and \ref{N_1_4}, we can intuitively build the expression of $N_1$ for a transaction with an arbitrary $n$. Specifically, we have:

\begin{equation}
N_1 =
\begin{pmatrix}
(-1)^n\frac{1}{\prod_{k=1}^n(p_k-1)}, (-1)^{n-1}\frac{1}{\prod_{k=2}^n(p_k-1)}, ... , (-1)^2\frac{1}{\prod_{k=n-1}^n(p_k-1)}, (-1)\frac{1}{p_n-1}
\end{pmatrix}.
\end{equation}

\noindent
Finally, denoting with $N_{1,i}$ the \textit{i-th} element of $N_1$  (we remark that it represents the average number of visits to state $i$ when the initial state is $O_1$), we have:
\begin{equation}
N_{1,i}=
(-1)^{n-(i-1)}\frac{1}{\prod_{k=i}^n(p_k-1)}.
\label{number_of_visits}
\end{equation}

\noindent
By using the vector $T$ and $t_C$, the average total transaction execution time $R$ can be calculated as follow:

\begin{equation}
R = N_1 \cdot T + t_C.
\label{transaction_response_time}
\end{equation}

\subsection{Lock holding times}

We note that a transaction, due to aborts and restarts, can acquire and release a lock on the same data item more than one time. Thus, the total time it keeps locked a data item is given by the sum of the durations of all locks acquired on the data item. Consequently, the average time a transaction keeps locked a data item written at the operation $O_i$ can be estimated as the total time that the transaction spends in all states $O_j$ with $j>i$, plus the time it spends in state $C$. Accordingly, we have:

\begin{equation}
l_i =\sum_{k=i+1}^n N_{1,k} \cdot T_k +t_c.
\label{lock_holding_time}
\end{equation}

\subsection{Lock conflict probability}
\label{lock_conflict_probability}
We note that the time $l_i$ during which a transaction keeps locked a data item is a fraction of the total transaction execution time. Then, denoting with $f_i$ the average value of this fraction of time, we have:

\begin{equation}
f_i =\frac{l_i}{R}.
\label{fraction}
\end{equation}

\noindent
Now, we  assume that that lock request arrival times of each transaction are independent from lock request arrival times of the other concurrent transactions. Thus, if a transaction $a$ executes a transactional operation on a data item $x$ at operation $O_i$, $f_i$ can be considered as the probability that a lock request by another concurrent transaction $b$ on the data item $x$ conflicts with the lock held by transaction $a$. 

\noindent
Since in the system there are $m$ concurrent threads executing transactions, when transaction executes a transactional operation on the data item $x$ the probability that no conflict occurs with the other $m-1$ concurrent transactions is equal to the probability that none of the other $m-1$ concurrent transactions is holding a lock on $x$. Thus, the conflict probability can be estimated as the sum of the times the other $m-1$ concurrent transactions keep locked $x$ divided the average execution times of transactions. 

\subsection{Transaction data access patterns}
Now we extend the model to capture the effects of transaction data access patterns of Scenario 1 and Scenario 2 described in section \ref{scenarios}. 

\begin{enumerate}

\item[-] \textbf{Scenario 1.} To model the transactional data access patters of Scenario 1, we partition the set of $d$ data items in $n$ data subsets, each one with size $s=d/n$. These subsets represent the different data tables.

\item[] \textit{\textbf{Case 1.1.}} We assume that a subset of threads execute transactions which access data subsets in the same order. Specifically, the first transactional operation accesses a random data item in first data subset $1$, the second transactional operation accesses a random data item in the data subset $2$, and so on. The remaining threads execute transactions which access data subsets in the inverse order, i.e. the first transactional operation accesses a random data item in the data subset $n$, the second transactional operation accesses a random data item in the data subset $n-1$, and so on. The sizes of the two subsets of threads are denoted with $m'$ and $m''$, respectively, having $m'+m''=m$. We note that a transactional operation accesses a given data item within each data subset with probability $1/s$.

\item[] \textit{\textbf{Case 1.2.}} In this case, we assume that all threads execute transactions which access the different data subsets in the same order. Specifically, the first transactional operation accesses a random data item in the data subset $1$, the second transactional operation accesses a random data item in the data subset $2$, and so on. Also in this case, a transactional operation accesses a given data item within each data subset with probability $1/s$.

\item[-] \textbf{Scenario 2.}  To model the transactional data access patters of Scenario 2, we consider as follows.

\item[] \textit{\textbf{Case 2.1.}} We assume that all threads execute transactions where all transactional operations access a random data item in the whole set of $m$ data items. Hence, a transactional operation accesses a given data item within each data subset with probability $1/m$. 

\item[] \textit{\textbf{Case 2.2.}} We assume that all threads execute transactions that access randomly selected data items in the whole set of $m$ data items, but data items are ordered in ascending way before the transaction execution.

\end{enumerate}

\noindent
For simplicity, we start from modelling Case 1.2. When a transaction accesses a data item at the transactional operation $O_i$, it accesses a random data item in the \textit{i-th} subset. Further, we know that all data items in the \textit{i-th} subset are accessed by all the other $m-1$ concurrent transactions at the transactional operation $O_i$. Consequently, all the other $m-1$ concurrent transactions access $x$ with probability $1/s$, and the average fraction of time they keep locked $x$ is $f_i$. Thus, based on our discussion at the end of Section \ref{lock_conflict_probability}, the lock conflict probability $p_i$ can be calculated as follows:

\begin{equation}
p_i=\frac{1}{s}(m-1)f_i.
\label{p_case11}
\end{equation}

\noindent
Now we consider Case 1.1, transactions executed by threads of the first subset execute the same data access pattern as in the first scenario. Accordingly, when a transaction that accesses a data item at the transactional operation $O_i$ the probability to conflict with one of the other $m'-1$ concurrent transactions executed by the first subset of threads, that we denote with $p_i'$, can be calculated as:

\begin{equation}
p_i'=\frac{1}{s} \left( m'-1 \right) f_i.
\label{p'_case12}
\end{equation}

\noindent
A transaction executed by a thread of the first subset can also conflict with transactions executed by threads of the second subset. Accordingly, when the transaction accesses a data item at the transactional operation $O_i$, the probability to conflict with one of the other $m''$ concurrent transactions executed by the second subset of threads, that we denote with $p_i''$, can be calculated as:

\begin{equation}
p_i''=\frac{1}{s}  m'' f_{n-i+1}.
\label{p''_case12}
\end{equation}

\noindent
Thus, the overall conflict probability when a transaction accesses a data item at the transactional operation $O_i$ is:

\begin{equation}
p_i=p_i'+p_i''.
\label{p_case12}
\end{equation}

\noindent
Now we consider Case 2.1. When a transaction accesses a data item at the transactional operation $O_i$, it accesses a random data item in the whole sub of $d$ data items. Since also all the other $m-1$ transactions access a random data item at each transactional operation, then the average fraction $f^{avg}$ of time a transaction keeps locked a data item is equal to:

\begin{equation}
f^{avg}=\sum_{k=1}^n \frac{f_k}{n}.
\label{favg_case21}
\end{equation}

\noindent
Thus, considering that all the other $m-1$ transactions execute $n$ transactional operations that access random data items, the lock conflict probability $p_i$ can be calculated as follows:

\begin{equation}
p_i=\frac{n}{d}\cdot f^{avg}\cdot(m-1).
\label{p_case21}
\end{equation}

\noindent
Finally, we consider Case 2.2. All transaction data access patterns in this case include $n$ data items randomly extracted from the set of $d$ data items, and then they are ordered in ascending way. Thus, we need to calculate for each transactional operation $O_i$, the probability that will be accessed each specific data items, i.e the probability $P_{i,x}$ that at operation $O_i$ a data item $x$ will be accessed. For simplicity, we consider that $x$ is the ordinal number of the data item. Thus, it is possible to show that (a proof is reported in the appendix of this article): 

\begin{equation}
P_{i,x}=\frac{\binom{x-1}{i-1} \cdot \binom{d-x}{n-i}}{\binom{d}{n}}.
\label{px_case22}
\end{equation}

\noindent
Consequently, the lock conflict probability $p_i$ can be calculated on basis of the probability to access the different data items at each transactional operation $O_i$. Specifically, assuming that a transaction accesses data item $x$ at transactional operation $O_i$, a conflict can occur with a concurrent transaction if the concurrent transaction accesses data item $x$ at one of its $n$ transactional operations. We note that the fraction of time the concurrent transaction keeps locked a date item changes depending on the transactional operation it is accessed. Thus, the average fraction of time can be calculated as follow:

\begin{equation}
f^{avg}_x=\sum_{k=1}^n {P_{k,x}} \cdot f_k.
\label{f_case22}
\end{equation} 

\noindent
Finally, the conflict probability $p_i$ can be estimated as follow:

\begin{equation}
p_i= \sum_{x=1}^d \left( {P_{i,x}} \cdot f^{avg}_x \right) \cdot (m-1).
\label{p_case22}
\end{equation}

\subsection{Solving the analytical model}
\label{solving}

The Analytical model can be solved using an iterative approach. Initially, values to variables $m$, $d$, $n$, $t_C$, to the vector $T$ and to $m'$ and $m''$ have to be assigned ($m'$ and $m''$ are required only for Case 1.2).  Then, we can assign the initial value $0$ to all conflict probability, i.e. $p_i=0$, with $1 \le i \le n$, and to $R$. Thus, for each iteration the following sequence of equations have to be solved:
\begin{itemize}
\item[-] Equations \ref{number_of_visits}, \ref{transaction_response_time},  \ref{lock_holding_time}, \ref{fraction}, and then
\item[-] Equations \ref{p_case11} (only for Case 1.1), otherwise
\item[-] Equations \ref{p'_case12}, \ref{p''_case12} and \ref{p_case12} (only for Case 1.2), otherwise
\item[-] Equations \ref{favg_case21} and \ref{p_case21} (only for Case 2.1), otherwise
\item[-] Equations \ref{px_case22}, \ref{f_case22} and  \ref{p_case22} (only for Case 2.2).
\end{itemize}

\noindent
At the end of each iteration, the calculated value of $R$ can be compared with 
the value of $R$ calculated during the previous iteration. If the absolute difference is less that an arbitrary value $\epsilon$, the iterations can stop. We note that $\epsilon$ expresses the precision that we expect in the evaluation of $R$. For example, if we use milliseconds as a time unit, and we set $\epsilon=0.001$, the expected error on $R$ is at most $2 \cdot 0.001=0.002$ milliseconds.

\bibliographystyle{unsrt}
\bibliography{bibliography}

\newpage

\section*{Appendix}

In this appendix we report a justification of Equation \ref{px_case22}. The associated problem statement can be defined as follows.  \\

\noindent
\textit{Problem statement}

We assume that $S$ is a set of $d$ elements, and that $\preceq$ a total order relation  on $S$. $n$ elements are randomly extracted from $S$ and then are ordered according to $\preceq$. $P_{i,x}$ denotes the probability that in $i$-th position of the ordered set of extracted elements there is the $x$-th element of the elements of $S$ ordered according to the $\preceq$. We have:

\begin{equation}
P_{i,x}=\frac{\binom{x-1}{i-1} \cdot \binom{n-x}{k-i}}{\binom{n}{k}}.
\label{prob_equation}
\end{equation}

The problem falls within the scope of the order statistics problems \cite{Order_Statistics_2003}, where $P_{i,x}$ corresponds to the probability density function of the $i$-th order statistic of the set of the $n$ randomly extracted elements. In the follow, we report an intuitive procedure for deriving the expression of $P_{i,x}$. \\

\noindent
\textit{Procedure}

For simplicity, we assume that $S$ is the set of the first $n$ natural numbers. The proof can be generalized because each total order relation on a set of $n$ elements induces a bijection with the first $n$ natural numbers. 

The possible subsets of $k$ elements extracted from $S$ are all the possible $k-$combinations of $S$. We denote with $C$ the set of all the ordered $k-$combinations of $S$. The cardinality of $C$ can be calculated as:

\begin{equation}
|C|=\binom{n}{k}.
\end{equation}

We have to evaluate the number of $k$-combinations of $C$ in which an element $x \in S$ appears in the $i$-th position. We denote with $C_{x,i} \subseteq C$ the subset of $k$-combinations in which $x$ appears in the $i$-th position. To achieve Equation \ref{prob_equation} we use an iterative reasoning, starting from $C_{x,1}$.

$C_{x,1}$ includes all $k$-combinations in which in the subsequent $k-1$ positions of position $1$ there is a $(k-1)$-combination of the other $n-x$  elements of $S$ that are greater than $x$. Thus the cardinality of $C_{x,1}$ is

\begin{equation}
|C_{x,1}|=\binom{n-x}{k-1}.
\end{equation}

$C_{x,2}$ includes all $k$-combinations in which:
\begin{itemize}
\item in position $1$ there is a $1$-combination of the $x-1$ elements of $S$ that are less than $x$, and
\item  in the subsequent $k-2$ positions of position $2$, there is a $(k-2)$-combination of the $n-x$ elements of $S$ that are greater than $x$.
\end{itemize}

Denoting with $C_{x,2}^p$ the set of $1$-combinations of the $x-1$ elements of $S$ that are less than $x$, and with $C_{x,2}^s$ the set of $(k-2)$-combinations of the $n-x$ elements of $S$ that are greater than $x$, we have

\begin{equation}
|C_{x,2}^p|=\binom{x-1}{1}
\end{equation}

\noindent
and 
\begin{equation}
|C_{x,2}^s|=\binom{n-x}{k-2},
\end{equation}

\noindent
thus we have: 
\begin{equation}
|C_{x,2}|=\binom{x-1}{1} \cdot \binom{n-x}{k-2}.
\end{equation}

$C_{x,3}$ includes all $k$-combinations in which:
\begin{itemize}
\item in positions $1$ and $2$ there is a $2$-combination of the $x-1$ elements of $S$ that are less than $x$, and
\item  in the subsequent $k-3$ positions of position $3$, there is a $(k-3)$-combination of the $n-x$ elements of $S$ that are greater than $x$.
\end{itemize}

Denoting with $C_{x,3}^p$ the set of $2$-combinations of the $x-1$ elements of $S$ that are less than $x$, and with $C_{x,3}^s$ the set of $(k-3)$-combinations of the $n-x$ elements of $S$ that are greater than $x$, we have

\begin{equation}
|C_{x,3}^p|=\binom{x-1}{2}
\end{equation}

\noindent
and
\begin{equation}
|C_{x,3}^s|=\binom{n-x}{k-3},
\end{equation}

\noindent
thus we have: 
\begin{equation}
|C_{x,3}|=\binom{x-1}{2} \cdot \binom{n-x}{k-3}.
\end{equation}

By incrementally applying this reasoning, we can conclude as follow. $C_{x,i}$ includes all $k$-combinations in which:
\begin{itemize}
\item in the $i-1$ previous positions of position $i$ there is a $(i-1)$-combination of the $x-1$ elements of $S$ that are less than $x$, and
\item in the subsequent $k-i$ positions of position $i$, there is a $(k-i)$-combination of the $n-x$ elements of $S$ that are greater than $x$.
\end{itemize}

Denoting with $C_{x,i}^p$ the set of $(i-1)$-combinations of the $x-1$ elements of $S$ that are less than $x$, and with $C_{x,3}^s$ the set of $(k-i)$-combinations of the $n-x$ elements of $S$ that are greater than $x$, we have

\begin{equation}
|C_{x,i}^p|=\binom{x-1}{i-1}
\end{equation}

\noindent
and 
\begin{equation}
|C_{x,i}^s|=\binom{n-x}{k-i},
\end{equation}

\noindent
thus we have: 
\begin{equation}
|C_{x,i}|=\binom{x-1}{i-1} \cdot \binom{n-x}{k-i}.
\end{equation}

In conclusion, the probability that $x \in S$ appears in the $i$-th position of the ordered set of the extracted elements can be calculated by dividing $|C_{x,i}|$ by $|C|$. Thus we have:

\begin{equation}
P_{i,x}=\frac{\binom{x-1}{i-1} \cdot \binom{n-x}{k-i}}{\binom{n}{k}}.
\end{equation}

\end{document}